\documentclass[aps,prb,twocolumn]{revtex4-2}
\usepackage{amssymb}
\usepackage{graphicx,amsmath}
\usepackage{color,colordvi}
\setcitestyle{numbers,square}

\newcommand{\bea}{\begin{eqnarray}}
\newcommand{\eea}{\end{eqnarray}}
\newcommand{\beq}{\begin{equation}}
\newcommand{\eeq}{\end{equation}}
\newcommand{\benu}{\begin{enumerate}}
\newcommand{\enu}{\end{enumerate}}
\newcommand{\la}{\langle}
\newcommand{\ra}{\rangle}
\newcommand{\al}{\alpha}

\newcommand{\om}{\omega}
\newcommand{\Om}{\Omega}
\newcommand{\ep}{\epsilon}

\newcommand{\ham}{\mathcal{H}}

\newcommand{\ptl}{\partial}

\newcommand{\bk}{{\bf k}}

\newcommand{\tpr}{t^{\prime}}

\begin{document}

\title{
Detection of squeezed phonons in pump-probe spectroscopy
}
\date{\today}

\author{Massil Lakehal,$^1$ Marco Schir\'{o},$^{2}$\footnote{On Leave from: Institut de Physique Th\'{e}orique,
Universit\'{e} Paris Saclay, CNRS, CEA, F-91191 Gif-sur-Yvette, France} Ilya M. Eremin,$^{3,4}$ and Indranil Paul$^1$}
\affiliation{
$^1$Laboratoire Mat\'{e}riaux et Ph\'{e}nom\`enes Quantiques, Universit\'{e} de Paris, CNRS, Paris 75013, France\\
$^2$JEIP, USR 3573 CNRS, Coll\`ege de France, PSL Research University, F-75321 Paris, France\\
$^3$Institut f\"{u}r Theoretische Physik III, Ruhr-Universit\"{a}t Bochum, D-44801 Bochum, Germany\\
$^4$National University of Science and Technology MISiS, 119049 Moscow, Russian Federation
}

\begin{abstract}
Robust engineering of phonon squeezed states in optically excited solids has emerged
as a promising tool to control and manipulate their properties. However, in contrast to quantum optical systems,
detection of phonon squeezing is subtle and elusive, and an important question is what constitutes an
unambiguous signature of it. The state of the art involves observing oscillations at twice the
phonon frequency in time resolved measurements of the out of equilibrium phonon fluctuation. Using Keldysh
formalism we show that such a signal is a necessary but not a sufficient signature of a squeezed phonon,
since we identify several mechanisms that do not involve squeezing and yet which produce similar oscillations.
We show that a reliable detection requires a time and frequency resolved measurement of the phonon spectral function.
\end{abstract}

\maketitle


\section{Introduction}

Recent advances in ultrafast pump-probe techniques have opened the possibility of controlling
quantum materials by light~\cite{reviews,okamoto07,fausti11,liu12}. This includes manipulating electronic
orders~\cite{Zong,mitrano16, Kogar,Buzzi,Li-Eckstein,Matthies-Eckstein}, as well as controlling
the lattice dynamics. Thus, it is well-known that femtosecond pumping can create a coherent
phonon, which manifests in the transient optical properties as oscillations with the phonon
frequency $\omega_0$~\cite{silvestri1985,yan1985,chen1990,zeiger1992,merlin1997,ishioka2009}, or induce changes
to the electronic structure~\cite{foerst11,Subedi,mankowsky14,lakehal19}.

An intriguing newer goal is to engineer phonons into non-classical states such as squeezed
states~\cite{hu96,drummond04} that can be described by a density matrix
$\rho_{sq} \equiv S^{\dagger} \rho_{th}S$, where $S= \exp[i r (b^2 + (b^{\dagger})^2)]$ with $(b, b^{\dagger})$
the phonon annihilation/creation operators, $r$ the squeezing parameter
and $\rho_{th}$ the thermal density matrix.  This can be promising for
light-induced superconductivity~\cite{knap16,kennes17}. Engineering of squeezed
states is also a major goal in quantum technologies with broad applications from metrology to quantum
information~\cite{Wollman,Clerk}.

A related important goal is to unambiguously \emph{detect} a squeezed state once they have been created.
Several pump-probe experiments have reported
signature of phonon squeezing, based on measurements of the fluctuation
$N_X(t) \equiv \la X(t)^2 \ra - \la X(t) \ra^2$
of the displacement $X$ of a phonon showing oscillations with
frequency 2$\om_0$~\cite{garrett1997a,garrett1997b,bartels2000,misochko2000,johnson09,zijlstra10,trigo2013,esposito15,benatti17}.
Indeed theoretically, when a harmonic oscillator is squeezed impulsively
at time $t=0$, it leads to $N_X(t) \sim \sin (2 \omega_0 t)$ at subsequent times $t>0$.
In contrast, since the average $\la X(t) \ra$ is subtracted in $N_X(t)$, per se, the fluctuations of
coherent phonons do not have such oscillatory noise, unless interaction effects induce it
via higher harmonic generation [see case (iii) below].
\begin{figure}[!!b]
\begin{center}
\includegraphics[width=8.7cm,trim=0 0 0 0]{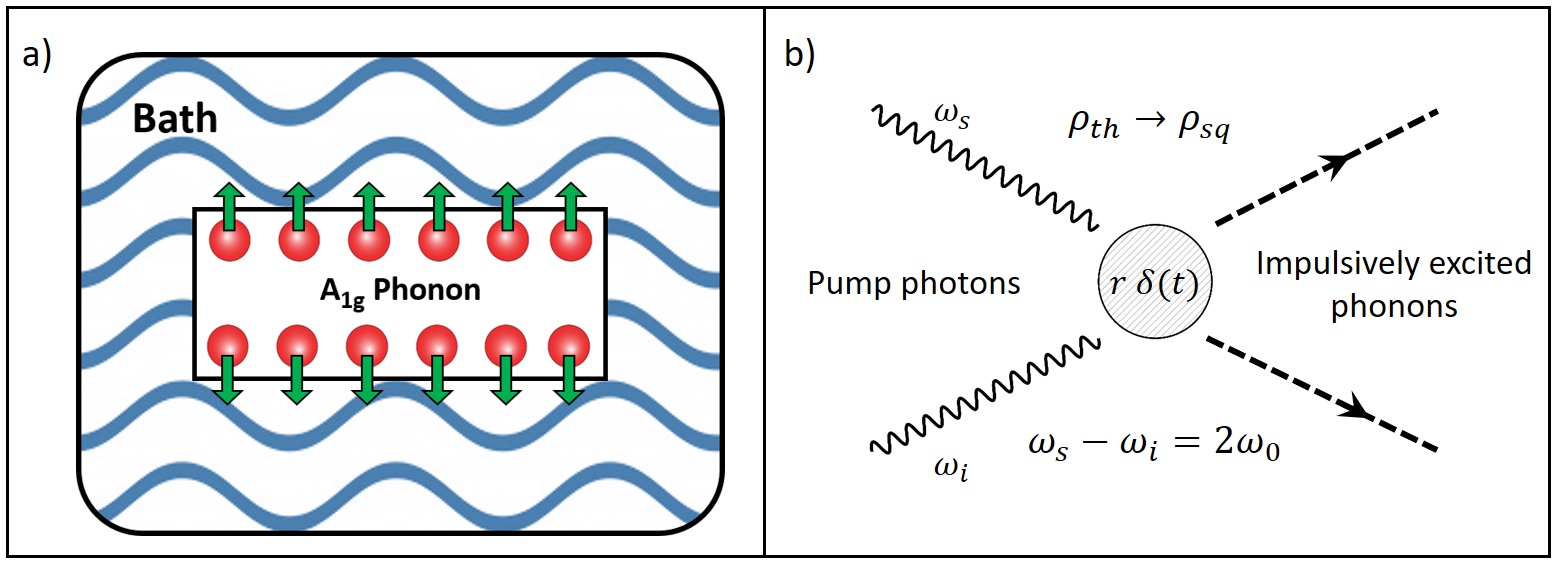}
\caption{
(a) Sketch of the system: a Raman phonon, frequency $\om_0$, in a bath.
(b) A pump-induced impulsive second order Raman process at time $t=0$ squeezing the phonon's thermal density matrix
$\rho_{th} \rightarrow \rho_{sq}$, with $r$ the squeeze parameter. $\omega_i$ ($\omega_s$) is incident (scattered)
pump photon frequency.}
\label{fig1}
\end{center}
\end{figure}

The purpose of the current work is twofold. (i) First, to bring into question the above state of the art
that relates $2\om_0$ noise oscillation with squeezing.
As we show below,
there are out of equilibrium mechanisms that do not involve phonon squeezing and yet which lead
to $ N_X(t) \sim \sin (2\omega_0 t)$.
Thus, by itself, such a signal
is not \emph{sufficient} to conclude having phonon squeezed state.
In fact, the main ingredient for $2\omega_0$ oscillatory signal appear to be the breaking
of time translation symmetry by a quench, which may or may not involve squeezing the phonon.
(ii) Second, we identify a more reliable method to detect a squeezed phonon state, which involves
both time- and frequency- resolved Raman measurement.

\section{Model}
We consider a Brillouin zone center Raman-active optical phonon with dimensionless canonical variables $(X, P)$,
coupled to a spinless fermionic bath with dispersion $\ep_{\bk}$, momentum $\bk$,
and described by operators $(c^{\dagger}_{\bk}, c_{\bk})$ [see Fig.~\ref{fig1}(a)].
The equilibrium Hamiltonian is
\beq
\label{eq:hamiltonian}
\ham = \frac{\hbar \om_0}{2} (X^2 + P^2) + \sum_{\bk}
\left( \ep_{\bk} + \frac{V X}{\sqrt{2N}} F_{\bk} \right) c^{\dagger}_{\bk} c_{\bk}.
\eeq
$V$ is the phonon-bath interaction energy, assumed to be small enough such that it can be treated perturbatively.
$F_{\bk}$ is a form factor that reflects the point group symmetry of the phonon, and $N$ is the total number of bath variables.
The role of the bath is simply to provide a finite inverse lifetime $\gamma$ to the phonon,
and to define a temperature $T$ for the
system. We assume $\om_0 > \gamma$.

\begin{figure}[!!t]
\begin{center}
\includegraphics[width=8.7cm,trim=0 0 0 0]{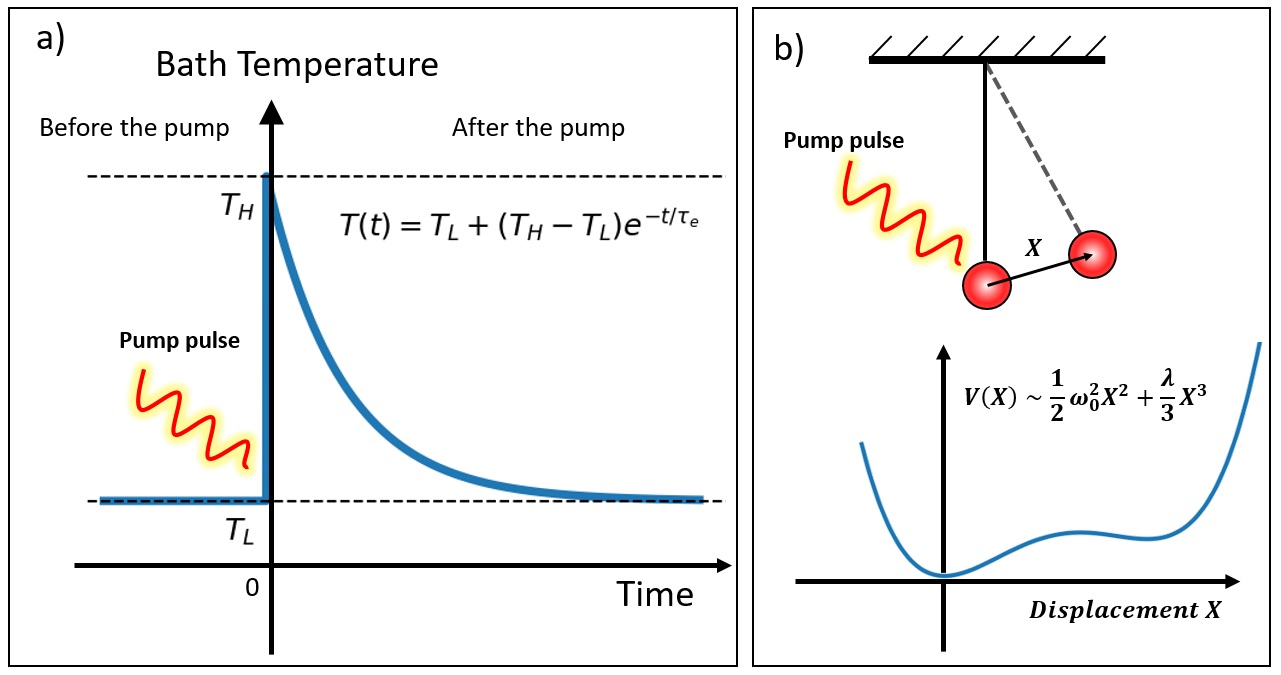}
\caption{
Two out-of-equilibrium processes that do not involve creating a squeezed phonon with density
matrix $\rho_{sq}$, but whose noise signal $N_X(t) \sim \sin (2 \om_0 t)$ has the same oscillatory
component as the squeezed case. (a) The bath temperature is quenched by the pump.
(b) A phonon, with a cubic anharmonic potential, is excited coherently by the pump.
}
\label{fig2}
\end{center}
\end{figure}
The system is subjected to a laser pump pulse of femtosecond duration at time $t=0$,
and we consider three different possible outcomes
of this out-of-equilibrium
perturbation that can manifest at a picosecond scale.
(i) First, we consider the outcome that the phonon is squeezed by the pump via an impulsive
resonant second order Raman
scattering process which leads to the generation/absorption of two phonon quanta [see Fig.~\ref{fig1}(b)]. The
perturbation is described by
$\delta \ham_{(i)} =r \delta(t) (X^2 - P^2)/2$, where $r$ is the
dimensionless squeezing factor.
(ii) Second, we consider the outcome that the pump leads to heating of the bath, such that the
bath temperature becomes time dependent. We model the thermal
quench as $T(t) = T_L$ for $t <0$, and $T(t) = T_L + (T_H - T_L) e^{-t/\tau_e}$, with $T_H \geq T_L$
[see Fig.~\ref{fig2}(a)].
Here $\tau_e$ is the thermal relaxation
timescale of the bath. In this case the bath density matrix is perturbed.
(iii) Third, we consider the outcome that the phonon,
assuming it has a symmetry-allowed cubic anharmonic potential,
is coherently excited by the pump pulse [see Fig.~\ref{fig2}(b)].
The generation mechanism of the coherent phonon, either via impulsive Raman scattering~\cite{merlin1997}
or via displacive excitation~\cite{zeiger1992}, is unimportant here.
This outcome is described by
$\delta \ham_{(iii)} = \hbar \om_0 f(t) X + \lambda X^3/3$, where $f(t)$ is the time-dependent force
that excites the phonon coherently, and
$\lambda$ is the energy scale of the cubic anharmonic potential. Importantly,
such a cubic anharmonicity is relevant in most systems that have been studied to date in the context of squeezing.
These include fully symmetric $A_g$-phonons in all systems, as well as
$E_g$-phonons in bismuth and $\al$-quartz with $D_3$ point group.
Note, for low enough pump fluence, the signatures of all these three outcomes depend linearly on the fluence.

Our goal is to study the fluctuation/noise $N_X(t)$ generated for each of the above three outcomes separately.
We work using the Keldysh formalism, where the phonon coordinate is defined on a two-branch
real-time contour, $X_{\pm}(t)$.
We integrate out the bath variables, treating the coupling $V$ to second order, and write the phonon
action in the more physical classical and quantum
basis $X_{cl/q}=\left(X_+\pm X_-\right)/\sqrt{2}$ as~\cite{KamenevBook,SchiroLeHur}
\beq
\label{eq:squeeze-action}
S[X] =  \int_{-\infty}^{\infty} d \tpr \int_{-\infty}^{\infty} dt \; X^T(\tpr) \hat{D}_{\rm inv} (\tpr, t) X(t),
\eeq
where $X^T(t) \equiv (X_{\rm cl}(t), X_{\rm q}(t))$, and the matrix
\[
\hat{D}_{\rm inv} (\tpr, t)  \equiv
\begin{bmatrix}
0 & D_A^{-1}(\tpr, t) \\
D_R^{-1}(\tpr, t) & - \Pi_K(\tpr,t)
\end{bmatrix}.
\]
$
D_{R/A}^{-1}(\tpr,t)=D_{0, R/A}^{-1}(\tpr,t)- \Pi_{R/A}(\tpr,t)
$
are the inverse retarded(advanced) phonon propagators and
$D_{0, R/A}^{-1}(\tpr,t)$ are those of a free phonon.
The self-energy contributions $\Pi_{R/A/K}(\tpr,t)$ from the fermionic bath are given by
\[
\Pi_R(\tpr,t) = \Pi_{{\rm cl}, {\rm q}}(\tpr,t)
\equiv -i V^2 \sum_{\bk} F_{\bk}^2 \la n_{\bk, {\rm cl}}(\tpr) n_{\bk, {\rm q}}(t) \ra_{S_{\rm b}^0},
\]
$\Pi_A(\tpr,t) = \Pi_{{\rm q}, {\rm cl}}(\tpr,t)$, and $\Pi_K(\tpr,t) = \Pi_{{\rm cl}, {\rm cl}}(\tpr,t)$.
$S_{\rm b}^0$ is the action of the bare bath, and
$n_{\bk, {\rm cl/q}} \equiv (c^{\dagger}_{\bk, {\rm cl}} c_{\bk, {\rm cl/q}}
+ c^{\dagger}_{\bk, {\rm q}} c_{\bk, {\rm q/cl}})/\sqrt{2}$.
From Eq.~(\ref{eq:squeeze-action}) we get
\beq
\label{eq:master-formula}
\la X_{\rm cl}(t)^2 \ra_{S[X]} = i \int_{-\infty}^{\infty}
dt^{\prime \prime} dt^{\prime} D_R(t, t^{\prime \prime}) D_R(t, t^{\prime})
\Pi_{K}(t^{\prime \prime}, t^{\prime}),
\eeq
which is the out-of-equilibrium generalization of the fluctuation-dissipation relation.
Our task is to obtain $D_R (\tpr, t)$ and $\Pi_K(\tpr,t)$, and from these to calculate the noise
$N_X(t)$ for the above three cases.

\section{Results}

Our main results are as follows. (1) We show that for
\emph{all} the three cases the noise signal has a $2\om_0$-oscillatory component with $N_X(t) \sim \sin (2 \omega_0 t)$.
Yet, only
in case (i) the signal is from a phonon squeezed state. Thus, a $2\omega_0$-oscillatory noise signal is \emph{not sufficient}
to conclude whether the phonon is in the squeezed state $\rho_{sq}$ or not. (2) By comparing the phonon spectral function
for the three cases,
we obtain a reliable method to detect a squeezed phonon state $\rho_{sq}$ which involves time-and
frequency-resolved Raman measurement.

\subsection{Case (i): Noise from squeezing}
As the bath itself is in equilibrium with temperature $T$,
$\Pi_K(\tpr,t)$ is a function of $(\tpr - t)$, and its Fourier transform $\Pi_{K, {\rm eq}}(\om)$ satisfies
standard fluctuation-dissipation theorem
$
\Pi_{K, {\rm eq}}(\om) = [ \Pi_{R, {\rm eq}}(\om) - \Pi_{A, {\rm eq}}(\om) ] \coth [\om/(2T)].
$
As we are interested in the long-time dynamics of the phonon, it is enough to expand $\Pi_{R, {\rm eq}}(\om)$
in frequency.
Since charge excitations are gapless in a good fermionic bath, we get
${\rm Im} \Pi_{R, {\rm eq}}(\om) \approx -\om \gamma/\om_0$, and
$
\Pi_{K, {\rm eq}}(\om) \approx - 2i \om \gamma \coth [\om/(2T)]/\om_0.
$
Next, due to $\delta \ham_{(i)}$ the squeezing perturbation, $D_R(\tpr, t)$ satisfies
\begin{align}
\left[ h(t) \ptl_t^2 + \dot{h}(t) \ptl_t + 2 \gamma \ptl_t + r \om_0 \delta(t) + \om_0^2 \right] D_R(\tpr, t) \nonumber \\
= - 2\om_0 \delta(\tpr - t), \nonumber
\end{align}
where $h(t) \equiv \om_0/[\om_0 - r \delta(t)]$, $\dot{h}(t) = \ptl_t h(t)$,
while $\gamma$ plays the role of phonon damping induced by the fermionic bath.
The solution of the above equation with the
initial conditions $D_R(t^+,t)=0$ and $\ptl_t D_R(t^+,t)= -2\om_0$ is described in detail
in Appendix~\ref{app-a}.
Here we simply quote the final answer.
Before that, for convenience, we introduce the notation
\[
\mathcal{F}_T[f(x_1, x_2, \cdots); x_i; k_i] \equiv \int_{-\infty}^{\infty} dx_i e^{i k_i x_i} f(x_1, x_2, \cdots).
\]
In terms of $\Delta_R(t, \om) \equiv \mathcal{F}_T [D_R (t, t - \tau); \tau; \om]$ we get
\beq
\label{eq:D-squeeze}
\Delta_R (t, \om) =\left[ 1 + \frac{i}{2\om_0} \theta(t) K(t, \om) e^{i \om t - \gamma t} \right] D_{R, {\rm eq}} (\om),
\eeq
where $ D_{R, {\rm eq}}(\om) \equiv 2\om_0/[\om^2 + 2i \gamma \om - \om_0^2]$ is the equilibrium propagator, and
$K(t, \om) \equiv A(t) (\om - \om_0 + i \gamma) e^{i \om_0 t} + \bar{A}(t) (\om + \om_0 + i \gamma) e^{-i \om_0 t}$.
The function $ A(t) \equiv \sinh r \cos (2\om_0 t) + i [ \cosh r - \sinh r \sin (2 \om_0 t) - 1]$ has information about
squeezing. Re-expressing Eq.~(\ref{eq:master-formula}) in terms of $\Delta_R(t, \om)$ and $\Pi_{K, {\rm eq}}(\om)$
we get~\cite{supplemental}
\beq
\label{eq:result-i}
N_X(t) = \frac{1}{2} e^{-2\gamma t} \sinh 2r [ \tanh r - \sin (2 \om_0 t)] \coth \frac{\om_0}{2T}.
\eeq
Thus, $N_X(t)$ has a $2\om_0$ oscillatory component.

\subsection{Case (ii): Noise from thermal quench of bath}
Here we assume that the effect of the pump on the bath can be encoded by a slowly
varying effective temperature, $T(s) = T_L + \theta(s) (T_H - T_L) e^{-s/\tau_e}$, for time
scales much longer than those relevant for the internal electronic dynamics. Thus, we disregard
the processes by which the bath itself thermalises to an effective time-dependent temperature, and we
study the consequences of such a pseudo-equilibrium environment on the phonon noise.

A first effect of the pseudo-equilibrium
is that the Keldysh self-energy $\Pi_K(\tpr,t)$, sensitive to the occupation of the bath modes,
loses time-translational invariance and acquires a dependence on the average time $s \equiv (\tpr + t)/2$ via
the temporally varying temperature.
It is convenient to define
$\Pi_K(s, \om) \equiv \mathcal{F}_T[\Pi_K(s + \tau/2, s - \tau/2); \tau; \om]$.
We expect that $\Pi_K(s, \om)$, in analogy with $\Pi_{K, {\rm eq}}(\om)$, satisfies the
fluctuation-dissipation relation with the
time-dependent temperature $T(s)$. This implies that
$
\Pi_K(s, \om) \approx - 2i \om \gamma \coth [\om/(2T(s))]/\om_0.
$
In principle, the retarded phonon propagator $D_R(\tpr,t)$ also acquires $s$-dependence
through the temperature dependence of damping $\gamma(s)$. However, to leading order
in the temperature quench $(T_H-T_L)$ this slow variation can be ignored,
and one can use the equilibrium form $ D_{R, {\rm eq}}(\om)$.
Note that, the above simplification does not affect the final conclusion qualitatively.

From the above considerations, it is simple to evaluate the fluctuations in terms of
$ \la X_{\rm cl}(\Om)^2 \ra_{S[X]}  \equiv \mathcal{F}_T [\la X_{\rm cl}(t)^2 \ra_{S[X]}; t; \Om]$ from the relation
\begin{align}
\label{eq:variance-Omega}
\la X_{\rm cl}(\Om)^2 \ra_{S[X]}
&= \frac{i}{4 \pi} \int_{- \infty}^{\infty} ds \int_{- \infty}^{\infty} d \om D_{R, {\rm eq}}(\om + \Om/2)
\nonumber \\
&\times D^{\ast}_{R, {\rm eq}}(\om - \Om/2) \Pi_K(s, \om) e^{i \Om s}.
\nonumber
\end{align}
We assume that $\gamma$ is the lowest energy scale and, in particular, $\gamma \ll T_L$. Then,
\[
\la X_{\rm cl}(\Om \sim \pm 2\om_0)^2 \ra_{S[X]} \propto \frac{1}{z_1 z_2}
\int_{- \infty}^{\infty} d s e^{i \Om s} \Pi_K(s, \om \rightarrow 0),
\]
where $z_{1,2} \equiv \Om/2 \pm \om_0 + i \gamma$. This implies
that $\la X_{\rm cl}(\Om)^2 \ra_{S[X]}$ has poles at $\Om = \pm 2 \om_0$, which in the
time domain translate as $\sin (2\om_0 t)$ oscillations.
Note, the origin of the these poles is an \emph{equilibrium} property of a damped oscillator. However,
in equilibrium time translation symmetry ensures that $\Pi_K(s, \om)$ does not depend upon the average time $s$.
In such a case the
$s$-integral above gives $\delta(\Om)$
and, therefore, in equilibrium the
finite-$\Om$ poles are ``inaccessible''. Keeping the $s$-dependence of $\Pi_K(s, \om)$ we get
\beq
\label{eq:result-ii}
N_X(t) = \frac{- \gamma (T_H - T_L)}{2\om_0^2} e^{-2\gamma t} \sin (2 \om_0 t) + \ldots,
\eeq
where the ellipsis imply non-oscillatory contributions.
Thus, we demonstrate that $N_X(t)$ can have $2\om_0$ oscillatory signal
simply due to pump induced thermal quench of the bath. In fact,
\emph{any} out-of-equilibrium perturbation of the bath that breaks time
translation symmetry will lead to a $2\om_0$ oscillatory noise signal.

\subsection{Case (iii): Noise from phonon with cubic anharmonicity coherently excited}
This is relevant for $A_g$ phonons in all systems, and also for
$E_g$ phonons in $D_3$ symmetric systems such as $\al$-quartz and bismuth.
The pump leads to $\la X_{\rm cl}(t) \ra_{S[X]} \equiv u(t) \neq 0$, where $u(t)$
describes the coherent excitation.
To study the fluctuations around the average we
expand the action in terms of $\delta X \equiv X - u(t)$. This gives
Eq.~(\ref{eq:squeeze-action}) with $X_{\rm cl}$ replaced by $\delta
X_{\rm cl}$, and with the propagator $D_R(\tpr, t)$ satisfying the
equation
\[
\left[ \ptl_t^2 + 2 \gamma \ptl_t + \om_0^2  + 4 \lambda \om_0 u(t) \right] D_R(\tpr, t)
= - 2\om_0 \delta(\tpr - t).
\]
In the above the crucial ingredient is the $\lambda u(t)$ term which
acts as a time dependent potential for the fluctuations.
We assume $\lambda/\om_0 \ll 1$, such that it is sufficient to solve the above equation to linear order in $\lambda$.
We expand $D_R(\tpr, t) = D_{R, {\rm eq}}(\tpr - t) + \lambda D_{R, 1}(\tpr, t)$ with
\[
D_{R, 1}(\tpr, t) = 2 \int_{-\infty}^{\infty} d s D_{R, {\rm eq}}(\tpr - s) u(s) D_{R, {\rm eq}}(s - t).
\]
On the other hand the bath is in equilibrium with temperature $T$, and therefore the Keldysh self-energy
is the one in equilibrium $\Pi_{K, {\rm eq}}(\om)$.
Ignoring the equilibrium contribution to noise we get
\begin{align}
& N_X(t)_{\rm non-eq} \nonumber \\
&= -i \lambda \int_{-\infty}^{\infty} d s  d s^{\prime}
D_{R, {\rm eq}}(t - s) D_{R, 1}(t, s^{\prime})
\Pi_{K, {\rm eq}}(s - s^{\prime})
\nonumber \\
& = 4 \lambda \coth (\frac{\om_0}{2T}) \int_0^t ds \; u(s) \sin [2\om_0(t-s)] e^{-2 \gamma(t-s)}.
\nonumber
\end{align}
In the last line the $s$-integral starts from $s>0$ because the coherent excitation is
triggered by the pump at $s=0$. This element of time translation symmetry breaking is crucial.
Irrespective of the details of the coherent motion $u(s)$, it is clear from the above that
there is an oscillatory $2\om_0$-signal in the noise. The precise form of $N_X(t)$ depend on
the force $f(s)$, which determines $u(s)$.
An impulsive excitation leads to $u(s) = u_0 \sin(\om_0 s) e^{-\gamma s}$, and
\beq
\label{eq:result-iii}
N_X(t)_{\rm non-eq} = -\frac{4 \lambda u_0 \coth[\om_0/(2T)]}{3 \om_0} e^{-2 \gamma t} \sin(2 \om_0 t) + \ldots,
\eeq
where the ellipsis imply terms not relevant for the current discussion. For a displacive
coherent excitation the sine signal is replaced by a cosine,
which reflects the property of coherent phonons that the phase of the oscillation is determined by the
generation mechanism~\cite{zeiger1992,merlin1997,ishioka2009}.
Thus, as in case (ii), we obtain oscillatory $2\om_0$- noise signal without having created
the density matrix $\rho_{sq}$.
However, in contrast to case (ii) which involves only incoherent excitations, in (iii) the $2\om_0$ signal is
built out of a coherent excitation at $\om_0$. Consequently, the latter is analogous to higher harmonic
generation.
\begin{figure}[!!t]
\begin{center}
\includegraphics[width=8cm,trim=0 0 0 0]{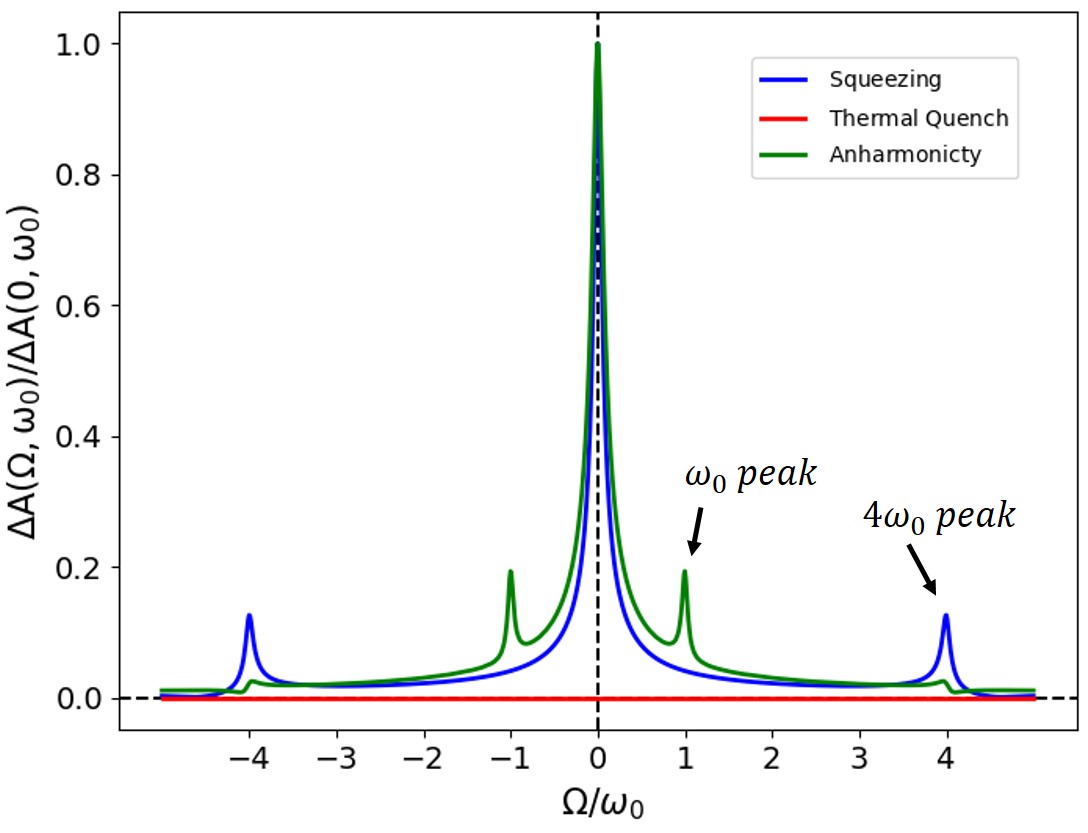}
\caption{
The time dependence of the spectral function peak $\Delta A(t, \omega_0) \equiv A(t, \omega_0) - A(0, \omega_0)$
(see Eq.~(\ref{eq:spectral}) for definition) and its Fourier transform $\Delta A(\Om, \omega_0)$ have distinct signatures
for the three cases studied here.
A distinguishing feature of squeezing is a \emph{single} peak in $|\Delta A(\Omega, \omega_0)|$ at $\Om = 4\om_0$.
}
\label{fig3}
\end{center}
\end{figure}

\subsection{Reliable signature of squeezed phonon}
Next, we identify a measurable quantity that can distinguish a squeezed phonon,  i.e., case (i) from (ii) and (iii).
Since the noise $N_X(t)$, which is a time resolved but frequency integrated quantity, has the same oscillatory property
for the three situations, a promising direction is to look for a quantity which is both time and frequency
resolved. With this intuition we define the time- and frequency- resolved spectral function
\beq
\label{eq:spectral}
A(t, \omega) \equiv {\rm Im} D_R (t, \omega),
\eeq
where $D_R (t, \omega) \equiv \mathcal{F}_T [D_R(t + \tau/2, t - \tau/2); \tau; \om]$
is the Wigner transform of the two-time retarded propagator.
Note, $D_R (t, \omega)$ is to be distinguished from $\Delta_R (t, \omega)$ defined earlier. In equilibrium,
$A(t, \omega)$ is $t$-independent and is peaked at $\om = \pm \om_0$. Thus, $A(t, \omega_0)$ describes how the
spectral peak varies with time $t$ after the pump.

We find that, indeed, the $t$-dependencies of $A(t, \omega_0)$ for the three cases are distinct,
since the phonon propagator $D_R (t, \omega)$ in the three cases are quite different
(for details see Appendix~\ref{app-b}).
In case (i) squeezing leads to $A(t, \omega_0)$ to oscillate with frequency 4$\om_0$ as a function of $t$
for the following reason. Since squeezing involves excitation of two phonon quanta,
$D_R(t + \tau/2, t - \tau/2) \sim \cos (2 \om_0 t) \theta (\tau - 2t)$. The constraint $\tau \geq 2t$ is crucial,
and follows from the fact that squeezing is an impulsive process occurring at time zero
[see Fig.~\ref{fig1}(b)], and so the two time arguments in
$D_R(t + \tau/2, t - \tau/2)$ must have opposite signs (see Eq.~(\ref{eq:a19}) in the Appendix).
Next, the Wigner transform involves multiplication of the phase $\exp(i \om \tau)$, and therefore,
imposing the constraint
leads, via phase accumulation, to  $A(t, \omega_0) \sim \cos (4 \om_0 t + \phi)$,
where $\phi$ is a phase that depends on details. Thus,
the $4\om_0$ oscillation is a consequence of impulsive excitation of two phonon quanta, which is the hallmark of
squeezing.
In case (ii)  $A(t, \omega_0)$ is practically $t$-independent, since the weak $t$-dependence of the phonon self-energy
induced by the electronic bath can be disregarded.
In case (iii)
$A(t, \omega_0)$ has oscillations at frequencies 4$\om_0$, but also at $\om_0$.
The latter is due to the fact that in this scenario the phonon is excited coherently.
This exemplifies that processes involving higher harmonic generation invariably will
have a signature at a frequency lower than $4\om_0$, and thus, they can be distinguished from squeezing.

These three distinct
$t$-dependencies can be conveniently expressed in terms of
$\Delta A(\Om, \om_0) \equiv \mathcal{F}_T[\Delta A(t, \om_0); t; \Om]$,
where $\Delta A(t, \om_0) \equiv A(t, \omega_0) - A(0, \omega_0)$ is the time dependent part of the
spectral function peak.
Thus, as shown in Fig.~\ref{fig3},
a distinguishing feature of squeezing is a \emph{single} peak in $|\Delta A(\Omega, \omega_0)|$ at $\Om = 4\om_0$.

Finally, we discuss how the time dependent spectral function $A(t, \omega)$ can be measured.
In equilibrium this is standard since the Raman
scattering cross-section is proportional to the correlation function, and from which the spectral function can
be deduced using fluctuation-dissipation relation. The issue is nontrivial in an out of equilibrium
situation since the standard fluctuation-dissipation relation is not valid. However, we can generalize the relation
in the following manner. According to the theory of time and frequency resolved Raman spectroscopy the scattering
intensity $I(t, \om) \propto D^> (t, \om)$, where $D^> (t, \om)$ is the Wigner transform of the two-time greater
function~\cite{Wang-Devereaux,Shvaika}.
Then, from standard Keldysh theory it follows that
$I(t, \om) - I(t, -\om) \propto A(t, \om)$ (see Appendix~\ref{app-c}).
In other words, the time dependent spectral function can be extracted from the difference of the Stokes
and anti-Stokes Raman intensities. In practice,
the probe cycle will require two pulses, one which is time-resolved and a second which is
frequency-resolved~\cite{Batignani,Dorfman}.

\section{Conclusion}
To summarize, we studied the fluctuations of a phonon coupled to a bath in a pump-probe setup. We argued
that the noise or the variance associated with the atomic displacement oscillates at twice the phonon frequency
$\om_0$, i.e. $N_X(t) \equiv \la X(t)^2 \ra - \la X(t) \ra^2 \sim \sin (2 \omega_0 t)$, due to pump-induced breaking
of time translation symmetry. Thus, such oscillations cannot be taken as proof of having prepared the phonon in a
squeezed state with density matrix $\rho_{sq}$. A reliable way to identify squeezed phonons involve a study of the
time-dependent spectral function
of the phonon which can be achieved by a time and frequency resolved Raman measurement.

\acknowledgments
We are thankful to A. Le Boit\'{e},
R. Chitra, M. Eckstein, D. Fausti, S. Florens, Y. Gallais, S. Johnson, Y. Laplace for insightful discussions.
M.~L. and I.~P. acknowledge financial support from ANR-DFG grant (ANR-15-CE30-0025).
I.~M.~E. was supported by the joint DFG-ANR Project (ER 463/8-1).
He also acknowledges partial support from
the Ministry of Science and Higher Education of the Russian Federation 
in the framework
of Increase Competitiveness Program of NUST MISiS Grant No. K2-2020-001.

\appendix

\section{Retarded Green's function of a squeezed phonon}
\label{app-a}

The retarded phonon Green's function of a squeezed phonon coupled to a thermal bath satisfies the equation of motion
\begin{align}
\label{SQ_DR_equation_Appendix}
&\frac{1}{2\omega_0}\left( h(t)\partial_{t}^{2}+\dot{h}(t)\partial_{t}+ 2\gamma \partial_{t}
+\omega_0^2+g(t)\omega_0   \right) D_{R}(t,t')
\nonumber\\
&=\delta(t-t'),
\end{align}
where $h(t) = \omega_0/(\omega_0-r g(t))$, and $g(t)$ describes the envelope of the pump-pulse.
We assume that the width of the pump-pulse is the smallest time scale of the problem, and hence,
that it can be well approximated by a Dirac distribution. It is, however, uncomfortable to deal
with the non-analyticity of the Dirac distribution in path integrals. Indeed, for a delta shaped
pump pulse centered at time $t=0$, the displacement $X_{cl/q}(t)$ and momentum $P_{cl/q}(t)(t)$
fields are not well defined at initial time $t=0$. In practice, however, the shape of a physical
pump-pulse is a smooth function with a finite width $\tau_{p}$, and the approximation with a Dirac
distribution is just a practical mathematical description. Therefore, to avoid complication,
we solve the problem for a general  pulse with a smooth envelope $g(t)$  centered at time $t=0$
and a finite width $\tau_p$, and only take the limit $\tau_{p}\rightarrow0$ for which
$g(t)\rightarrow r\delta{(t)}$ at the end of the calculation.

The Dirac delta function $\delta(t-t')$ is vanishing for $t\neq t'$, hence, for time
$t\neq t'$ the Green's function satisfies the homogeneous equation
\begin{equation}\label{SQ_DR_equation}
\bigg( h(t)\partial_{t}^{2}+h'(t)\partial_{t}+ 2\gamma \partial_{t}+\omega_0^2+g(t)\omega_0   \bigg) D_{R}(t,t')=0.
\end{equation}
The retarded Green's function has a causal structure, i.e. it  vanishes for $t<t'$,
and satisfies a second order differential equation which boundary condition are given by
the definition of the retarded Green's function at equal time
\begin{equation}\label{Fluctuations_SQ_boundary1}
D_{R}(t^{+},t)=-i\langle \left[ \hat{ X}(t^{+}),\hat{X}(t) \right] \rangle = 0,
\end{equation}
and the  jump condition imposed by the delta function
\begin{equation}\label{Fluctuations_SQ_boundary2}
\partial_{t} D_{R}(t^{+},t) = - 2\omega_0 ,
\end{equation}
where  $t^{+}=t+0^{+}$.

We  replace the equation of the retarded Green's function Eq. (\ref{SQ_DR_equation_Appendix})
by  a set of two coupled first order equations and write
\begin{subequations}\label{Fluctuations_SQ_foe}
\begin{align}
\partial_{t}D_{R}(t,s)  &=  \left[\omega_0- g(t) \right]K(t,s), \\
\partial_{t}K(t,s) & =  - \left[\omega_0+ g(t) \right]D_{R}(t,s)-2 \frac{\gamma }{\omega_0}\partial_t D_{R}(t,s),
\end{align}
\end{subequations}
where $K(t,s)$ is a function that we introduce as a mathematical tool. Notice that for $g(t)=0$,
the equation of the retarded Green's function Eq.~(\ref{Fluctuations_SQ_foe}) is that of a damped
harmonic oscillator. Therefore, we propose the following ansatz for the solution
\begin{subequations}
\begin{align}
D_{R}(t,s) &= X_{0}(t,s) \cos(\Omega t)e^{-\gamma t} \nonumber \\
&+ \frac{1}{\Omega}\left[ \omega_0P_{0}(t,s)
+ \gamma X_{0}(t,s) \right] \sin(\Omega t) e^{-\gamma t},\label{Fluctuations_SQ_DRform}  \\
K(t,s) &=P_{0}(t,s) \cos(\Omega t) e^{-\gamma t} \nonumber \\
&- \frac{1}{\Omega}\left[ \omega_0X_{0}(t,s)
+ \gamma P_{0}(t,s) \right] \sin(\Omega t) e^{-\gamma t},
\end{align}
\end{subequations}
where $\Omega^{2}=\omega_0^2-\gamma^{2}$. This form of the solution ensures that for constant $X_{0}(t,s)$
and $P_{0}(t,s)$, the functions $D_{R}(t,s)$ and $K(t,s)$ satisfy the equation of motion of a damped
harmonic oscillator. The fact that the phonon is a well defined excitation implies that $\gamma \ll \omega_0$.
Thus, we write the equations of motion for $X_{0}(t,s)$ and $P_{0}(t,s)$ in this limit and find that
\begin{equation}\label{Fluctuations_SQ_DE}
\begin{split}
\begin{bmatrix} \partial_t X_{0}(t,s) \\  \partial_t P_{0}(t,s)  \end{bmatrix}  &=-g(t)\begin{bmatrix}
-\sin{\left(2\Omega t\right)} &\cos{\left(2\Omega t\right)}  \\\cos{\left(2\Omega t\right)}
&  \sin{\left(2\Omega t\right)} \end{bmatrix}\begin{bmatrix} X_{0}(t,s)  \\  P_{0}(t,s)  \end{bmatrix}.
\end{split}
\end{equation}
We take the limit where the width of the pump-pulse goes to zero $\tau_p\rightarrow 0$ and recover
the Dirac delta function  $g(t) \rightarrow r \delta(t)$. Using the properties of the Dirac distribution, we have that
\begin{subequations}
\begin{align}
\sin(\Omega t) g(t) &= \sin(0)g(t)=0, \\
\cos(\Omega t) g(t) &= \cos(0) g(t)=g(t),
\end{align}
\end{subequations}
hence Eq.~(\ref{Fluctuations_SQ_DE}) further simplifies, and we obtain
\begin{equation}
\begin{split}
\begin{bmatrix} \partial_t X_{0}(t,s) \\  \partial_t P_{0}(t,s)  \end{bmatrix} &=-g(t)
\begin{bmatrix} 0 &1 \\1  &  0 \end{bmatrix}\begin{bmatrix} X_{0}(t,s)  \\  P_{0}(t,s)  \end{bmatrix}.
\end{split}
\end{equation}

The solution of this equation of motion is straightforward, and after some algebraic manipulations we find that
\begin{subequations}
\begin{align}
X_{0}(t,s) &=  \mathrm{ch}[F(t,s)] K_{1}(s) - \mathrm{sh}[F(t,s)]K_{2}(s),\\
P_{0}(t,s) &= \mathrm{ch}[F(t,s)] K_{2}(s) - \mathrm{sh}[F(t,s)]K_{1}(s),
\end{align}
\end{subequations}
with
\begin{equation}\label{SQ_Fdef}
F(t,s)=\int_{s}^{t}g(t')dt',
\end{equation}
where $\mathrm{ch}(x)$ and $\mathrm{sh}(x)$ denote the cosine and sine hyperbolic functions, respectively.
$K_{1}(s)$ and $K_{2}(s)$ are arbitrary functions  independent of time $t$,  that are to be calculated using
the boundary conditions, see Eq.~(\ref{Fluctuations_SQ_boundary1}) and Eq.~(\ref{Fluctuations_SQ_boundary2}).
We take the limit $\gamma\ll\omega_0$, and write the propagator in a more convenient form
\begin{align}
\label{SQ_DR_exp1}
D_{R}(t,s) & = \left( \mathrm{ch}[F(t,s)] \sin{\left[\Omega (t-s) -\phi(s)\right]}
\right. \nonumber \\
&- \left. \mathrm{sh}[F(t,s)]
\cos{\left[\Omega (t+s) + \phi(s)\right]} \right)A(s)e^{-\gamma (t-s)},
\end{align}
where we replaced the functions K1(s) and K2(s) by two other unknown functions $A(s)$ and $\phi(s)$ defined as
\begin{subequations}
\begin{align}
K_{1}(s) &=-A(s)\sin{\big(\Omega s + \phi(s)\big)}e^{\gamma s}, \\
K_{2}(s) &=A(s)\cos{\big(\Omega s + \phi(s)\big)}e^{\gamma s}.
\end{align}
\end{subequations}
The functions $A(s)$  and $\phi(s)$ can be evaluated using the boundary conditions. We use the continuity
condition  Eq.~(\ref{Fluctuations_SQ_boundary1}) of the retarded Green's function
\begin{equation}
\sin{\left[-\phi(t)\right]} A(t) = 0 ,\\
\end{equation}
together with the jump condition Eq.~(\ref{Fluctuations_SQ_boundary1}) at time $t\neq 0$
\begin{equation}
\begin{split}\label{}
A(t) \bigg(\Omega-\delta(t)\cos{(2 \Omega t)} \bigg) = -2\omega_0,
\end{split}
\end{equation} and find that
\begin{subequations}
\begin{align}
\phi(s) &= 0, \\
A(s) &= -2,
\end{align}
\end{subequations}
which gives for the retarded Green's function
\begin{align}
D_{R}(t,s)&= -2\theta(t-s) \left( \mathrm{ch}[F(t,s)] \sin{\left[\Omega (t-s)\right]}
\right. \nonumber \\
&- \left. \mathrm{sh}[F(t,s)] \cos{\left[\Omega(t+s) \right]} \right)e^{-\gamma (t-s) }.
\end{align}

In the limit of delta shaped pump-pulse  $g(t)=r\delta(t) $, we have that
\begin{subequations}
\begin{align}
 F(t,s) &= r\int^{t}_{s}\delta(t')dt'=r,              &              &\mathrm{sign}(t)\neq \mathrm{sign}(s)\\
 F(t,s) &=  r\int^{t}_{s}\delta(t')dt'=0, & & \mathrm{sign}(t)=  \mathrm{sign}(s)
\end{align}
\end{subequations}
with $\mathrm{sign}(t)$ the sign function. Thus, the retarded phonon Green's function is defined piece-wise and reads
\begin{equation}
\label{eq:a19}
D_{R}(t,s)=\
\begin{cases}
 D^{R}_{eq}(t,s),     &      \qquad         \rm{sign}(t) =\rm{sign}(s), \\
 D^{R}_{sq}(t,s),     &        \qquad    \rm{sign}(t) \neq\rm{sign}(s),
\end{cases}
\end{equation}
with
\begin{subequations}\label{A_SQ_propagator}
\begin{align}
 D^{R}_{eq}(t,s) &= -2 \theta(t-s)  e^{-\gamma(t-s)}\sin{\left[\Omega(t-s)\right] },          \\
D^{R}_{sq}(t,s)  &=  -2 e^{-\gamma (t-s) } \theta(t-s) \left( \mathrm{ch}(r) \sin{\left[\Omega (t-s)\right]} \right.
\nonumber \\
&-  \left. \mathrm{sh}(r) \cos{\left[\Omega(t+s) \right]} \right) ,
\end{align}
\end{subequations}
where $D^{R}_{eq}(t,s)$ and $D^{R}_{sq}(t,s)$ stand for the equilibrium and squeezed
retarded Green's function.
The fact that the squeezed propagator  is not a function of the the time difference
$D^{R}_{sq}(t,s)\neq D^{R}_{sq}(t-s)$ is a consequence of breaking time translational symmetry.

We now evaluate the function $\Delta_{R}(t,\omega)$ discussed in the main text and defined as
\begin{equation}\label{A_SQ_propagator_FTRSF}
\Delta_{R}(t,\omega) = \int \Delta_{R}(t,t-\tau)e^{i\omega \tau} d\tau.
\end{equation}
In time domain, the phonon propagators  $D_{R}(t_1,t_2)$ of the squeezed phonon is defined
piece-wise, see Eq.~(\ref{A_SQ_propagator}). Thus, we split the integral Eq.~(\ref{A_SQ_propagator_FTRSF})
over the two time domains and write
\begin{align}
\label{SQ_propagator_TRSF2}
\Delta_{R}(t,\omega)
&=\int\limits_{-\infty}^{\infty}  D_{eq}^{R}(t,t-\tau)   e^{i\omega \tau}d\tau
\nonumber\\
&+  \int\limits_{t}^{\infty}
\left( D_{sq}^{R}(t,t-\tau)-D_{eq}^{R}(t,t-\tau) \right)   e^{i\omega \tau}d\tau,
\end{align}
where we added and subtracted  $\int_{t}^{\infty}  D_{eq}^{R}(t,t-\tau)   e^{i\omega \tau}d\tau$
to extend the limits of the first integral to $+\infty$. The first term of the integral is the equilibrium
contribution and is time independent. The second term is the out-of-equilibrium contribution to the retarded
propagator $\Delta_{R}(t,\omega)$, and vanishes for a vanishing squeezing parameter $r=0$ or for negative
times $t<0$.  For positive times after the pump-pulse $t>0$, we have
\begin{equation}
\begin{split}\small
  &I(t,\omega) \equiv\int_{t}^{\infty}  \left( D_{sq}^{R}(t,t-\tau)  -D_{eq}^{R}(\tau) \right)   e^{i\omega \tau}d\tau \\
&=- 2\int_{t}^{\infty}  \bigg( g_{2}(t) \sin{\left(\Omega \tau\right)}- g_{1}(t)\cos{\left(\Omega\tau \right)} \bigg)
e^{(i\omega-\gamma)\tau} d\tau, \\
\end{split}
\end{equation}
where the complex functions $g_{1}(t)$ and $g_{2}(t)$ are defined as
\begin{subequations}\label{SQ_defA}
\begin{align}
g_{1}(t) &=\mathrm{sh}{(r)}\cos{(2\Omega t)}, \\
g_{2}(t) &=\mathrm{ch}{(r)}-\mathrm{sh}{(r)}\sin{(2\Omega t)}-1,
\end{align}
\end{subequations}
with $\Omega=\sqrt{\omega_0^2-\gamma^2}$. We define the complex function $A(t)=g_1(t)+ig_2(t)$ and express the
non equilibrium part of the retarded Green's function $I(\omega,t)$ in a compact exponential form
\begin{equation}
\begin{split}
I(t,\omega) &= \int_{t}^{\infty}  \bigg( A(t) e^{i\Omega \tau}+\bar{A}(t)e^{-i\Omega\tau}\bigg)
e^{(i\omega-\gamma)\tau} d\tau,\\
&=  \frac{i}{2\Omega}  \left(   A(t)\big( \omega-\Omega+i\gamma \big)e^{i( \omega +\Omega +i\gamma) t}
\right. \nonumber \\
&+ \left.
\bar{A}(t)\big( \omega+\Omega+i\gamma \big)e^{i( \omega -\Omega +i\gamma) t} \right)   D_{eq}^{R}(\omega),
\end{split}
\end{equation}
where $D_{eq}^{R}(\omega)$ denotes the equilibrium retarded Green's function. Henceforth, we take the
limit $\gamma\ll\omega_0$ for which $\Omega \approx \omega_0$, and we write the function $\Delta_{R}(t,\omega) $ as
\begin{equation}\label{SQ_DR_FTRS_exp}
\begin{split}
\Delta_{R}(t,\omega) &=\bigg( 1  +\frac{i}{2\omega_0} \theta(t) e^{-\gamma t} K(t,\omega)
e^{i\omega t} \bigg) D_{eq}^{R}(\omega),
\end{split}
\end{equation}
so that
\begin{subequations}
\begin{align}
 D^{R}_{eq}(\omega) &=\frac{2\omega_0}{\omega^2+2i\gamma\omega-\omega_0^2}, \\
 K(t,\omega) & =   A(t)\big( \omega-\omega_0+i\gamma \big)e^{i\omega_0  t}
 \nonumber\\
 &+\bar{A}(t)
\big( \omega +\omega_0+i\gamma \big)e^{-i \omega_0 t}.\label{SQ_Kdef}
 \end{align}
\end{subequations}
 This gives Eq.~(4) in the main text. The out of equilibrium part of the retarded Green's function,
as defined in Eq.~(\ref{A_SQ_propagator_FTRSF}), has a $2\omega_0$ oscillatory component. This follows
from the definition of the function $A(t)$, whose real and imaginary part oscillate as $\cos(2\omega_0 t)$
and $\sin(2\omega_0 t)$, respectively.

We insert the Fourier transform of the retarded Green's function $\Delta_{R}(t,\omega)$ in the expression of
the equal time correlation function   and write for time $t>0$
\begin{equation}
\begin{split}
\langle X^{2}_{cl}(t)\rangle &=\frac{i}{2\pi}\int\limits^{+\infty}_{-\infty} \Delta_{R}(t,\omega)D^{R}(t,-\omega)
\Pi^{K}(\omega)d\omega,\\
&=\frac{1}{2\pi}\int\limits^{+\infty}_{-\infty} \bigg( 1  - y(\omega,t)\bigg)D_{eq}^{K}(\omega)d\omega,\\
\end{split}
\end{equation}
where $y(\omega,t)$ is the non equilibrium part of the Keldysh Green's function and is given by
\begin{align}
y(\omega,t) &\equiv \frac{1}{4\omega_0^2}  e^{-2\gamma t}K(t,\omega) K(t,-\omega)
\nonumber\\
&-\frac{i}{2\omega_0} e^{-\gamma t}  \bigg( K(t,-\omega)e^{-i\omega t}+K(t,\omega)e^{i\omega t}  \bigg).
\end{align}
We recall that the equilibrium Keldysh Green's function is given  by
\begin{equation}
\begin{split}
D^{K}_{eq}(\omega) &= \frac{  8 \gamma\omega_0\omega \coth{\left[ \omega/2T\right]} }
{\left(  \left[\omega-\omega_0\right]^2+\gamma^2\right)\left( \left[\omega+\omega_0\right]^2+\gamma^2\right)}
\nonumber\\
&=  \frac{  2\gamma  \coth{\left[ \omega/2T\right]} }{ (\omega-\omega_0)^2+\gamma^2}-\frac{2\gamma
\coth{\left[ \omega/2T\right]}}{ (\omega+\omega_0)^2+\gamma^2},
\end{split}
\end{equation}
where we see that it is peaked at $\omega=\pm\omega_0$ with a width $\gamma$. Therefore, in
the limit where $\gamma\ll \omega_0$,  the $\coth{\left[ \omega/2T\right]}$ is a slow function of
frequency and the Keldysh Green's function can be approximated by
\begin{equation}
\begin{split}
D^{K}_{eq}(\omega) &\approx\frac{2\gamma  \coth{\left[ \omega_{0}/2T\right]} }
{ (\omega-\omega_{0})^2+\gamma^2}+\frac{2\gamma \coth{\left[ \omega_{0}/2T\right]}}{ (\omega+\omega_{0})^2+\gamma^2},
\end{split}
\end{equation}
where we used $\coth{\left[ -\omega_{0}/2T\right]}=-\coth{\left[ \omega_{0}/2T\right]}$.
We replace in the expression of the equal time correlation function of the classical field and obtain
\begin{equation}\label{Fluctuatios_SQ_var_final}
\begin{split}
\langle X^{2}_{cl}(t)\rangle  &=\frac{2}{\pi}\coth{\left[ \omega_{0}/2T\right]}
\int\limits^{+\infty}_{-\infty}\frac{\gamma   }{ (\omega-\omega_{0})^2+\gamma^2}
\\ &\times\left[ 1  -\frac{1}{4\omega_0^2}  e^{-2\gamma t}K(t,\omega) K(t,-\omega) \right.
\\
+& \left. \frac{i}{2\omega_0} e^{-\gamma t}  \bigg( K(t,-\omega)e^{-i\omega t}+K(t,\omega)e^{i\omega t}  \bigg) \right]d\omega.
\end{split}
\end{equation}

We integrate over the frequency $\omega$ and  obtain for the equal time correlation function
\begin{widetext}
\begin{equation}
\begin{split}
\langle X^{2}_{cl}(t)\rangle  = 2\coth{\left[ \omega_{0}/2T\right]}& \bigg[  1   -\frac{1 }{4\omega_0^2}
e^{-2\gamma t}K(t,\omega_{0}+i\gamma) K(t,-\omega_{0}-i\gamma)    \\&+\frac{i }{2\omega_0} e^{-2 \gamma t}
\bigg( K(t,-\omega_{0}-i\gamma)e^{-i\omega_0 t}+K(t,\omega_0+i\gamma )e^{i\omega_0 t}  \bigg) \bigg].  \\
\end{split}
\end{equation}
\end{widetext}
We take the limit $\gamma\ll\omega_0$, and write the correlation function as
\begin{equation}
\begin{split}
\langle X^{2}_{cl}(t)\rangle  &=2\coth{\left[ \omega_{0}/2T\right]} \left[ 1  + e^{-2\gamma t}\bar{A}(t)A(t)
\right. \\
&+ \left. i  e^{-2 \gamma t}  \left(\bar{A}(t)  - A(t)\right) \right].\\
\end{split}
\end{equation}
We replace the complex function $A(t)$ by its expression in Eq.~(\ref{SQ_defA}) and obtain for the variance after simplification
\begin{equation}\label{SQ_variance_final}
\begin{split}
\Delta^{2} X(t)  &= \coth{\left[ \omega_{0}/2T\right]} \left[ 1  + e^{-2 \gamma t}
\left( \mathrm{ch}(2r)-1\right) \right. \\
&- \left. \mathrm{sh}{(2r)}\sin{\big( 2\omega_0 t \big)}e^{-2\gamma t}\right],\\
\end{split}
\end{equation}
where $r$ is the squeezing parameters. The variance of the atomic displacement oscillates at
twice the frequency of the mode $2\omega_0$.   This is Eq. (5) in the main text.

\section{Time resolved spectral function}
\label{app-b}

Here, we discuss the calculation details of the time resolved spectral function defined as
\begin{equation}
\begin{split}
A(t,\omega) &\equiv \mathrm{Im} D^{R}(t,\omega),
\end{split}
\end{equation}
where
\begin{equation}
 D^{R}(t,\omega) = \int D^{R}(t+\tau/2,t-\tau/2) e^{i\omega \tau }d\tau,
\end{equation}
 is the Wigner transform of the retarded Green's function. We derive the expression of
$A(t,\omega)$ in the two scenarios discussed in the main text, namely  a squeezed phonon and a phonon,
with a cubic anharmonic potential,  excited coherently by the pump. We show that the time resolved
spectral is qualitatively different for this two case, and can therefore be used to detect phonon squeezed states.

\emph{Case (i)}: Let us first discuss the signature of  a squeezed phonon  (case $ (i)$  of the main text)
on the spectral function $A(t,\omega)$. The out of equilibrium part of the Wigner transform is defined as
\begin{equation}\label{A_DR_Wigner}
\begin{split}
&\Delta D^{R}(t,\omega) \equiv D^{R}(t,\omega) -D_{eq}^{R}(\omega), \\
&= \int\limits_{-\infty}^{+\infty} \bigg(  D^{R}(t+\tau/2,t-\tau/2) -D_{eq}^{R}(\tau)\bigg)e^{i\omega \tau} d\tau,
\end{split}
 \end{equation}
where $D^{R}_{eq}(\omega)$ is the equilibrium retarded Green's function. In equilibrium, the Wigner
transform coincides with its Fourier transform, hence, the retarded Green's function $ D^{R}(t+\tau/2,t-\tau/2)$
of a squeezed phonon is different from equilibrium only for times where
$\mathrm{Sign}(t+\tau/2) \neq \mathrm{Sign}(t-\tau/2)$. Therefore, the integral in Eq.(\ref{A_DR_Wigner})
is non vanishing only for time $\tau>2t$ where it can be written as
\begin{equation}\label{A_DR_Wigner2}
\begin{split}
\Delta D^{R}(t,\omega)  &= \int\limits_{2t}^{\infty} \left[  D^{R}_{sq}(t+\frac{\tau}{2},t-\frac{\tau}{2})
-D_{eq}^{R}(\tau)\right] e^{i\omega \tau} d\tau,\\
  &= -2 \big(\mathrm{ch}(r)-1\big)\int\limits_{2t}^{\infty} \sin(\omega_0\tau)
  e^{-\gamma \tau}
e^{i\omega \tau} d\tau \\
&+2 \mathrm{sh}(r)\cos(2\omega_0 t)\int\limits_{2t}^{\infty}
e^{-\gamma \tau} e^{i\omega \tau} d\tau.
\end{split}
 \end{equation}
We get
\begin{equation}
\begin{split}
&\Delta D^{R}(t,\omega)= +2 \mathrm{sh}(r)\cos(2\omega_0 t)
\int\limits_{2t}^{\infty}e^{-\gamma \tau} e^{i\omega \tau}d\tau\\
&+ i\big(\mathrm{ch}(r)-1\big)\int\limits_{2t}^{\infty} \left[e^{i(\omega+\omega_0+i\gamma )\tau}
-e^{i(\omega-\omega_0+i\gamma )\tau}\right] d\tau.
\end{split}
 \end{equation}
We integrate over time $\tau$ and find
\begin{widetext}
\begin{equation}
\begin{split}
\Delta D^{R}(t,\omega)
  =  \big(1 - \mathrm{ch}(r)\big)e^{-2\gamma t}\left[  \frac{e^{ 2 i
(\omega+\omega_0)  t} }{\omega+\omega_0+i\gamma}
-  \frac{e^{2 i(\omega - \omega_0)t} }{\omega-\omega_0+i\gamma} \right]
+2i \mathrm{sh}(r)\cos(2\omega_0 t)e^{-2\gamma t} \frac{e^{ 2 i\omega t}}{\omega+i\gamma}.
\end{split}
 \end{equation}
 \end{widetext}
 Thus the spectral function $A(t, \omega_0)$ evaluated at frequency $\omega_0$ has a $4\omega_0$ oscillatory
component in time domain, coming from the first and third terms above.

\emph{ Case (iii)}: We now calculate the spectral function of a phonon, with a cubic anharmonic potential,
excited coherently by the pump  (case $ (iii)$  of the main text). The first order correction of the
retarded Green's function, discussed in the main text, is given by
\begin{equation}
D_{R,1}(t,t') =2\int D_{R,eq}(t-s)D_{R,eq}(t'-s)u(s)ds.
\end{equation}
We Wigner transform the retarded Green's function and write
\begin{align}
D_{R,1}(t,\omega) &= 2\int D_{R,eq}(t+\tau/2 -s) \nonumber\\
&\times D_{R,eq}(t-\tau/2-s)u(s)e^{i\omega\tau}d\tau ds.
\end{align}
For simplicity, we discuss the case of an impulsive stimulation for which the retarded Green's function
and the average atomic displacement $u(t)$ are given by
\begin{equation}
\begin{split}
D_{R,eq}(t,t') &= -2\theta(t-t') \mathrm{ sin}\left[\omega_0(t-t')\right]e^{-\gamma(t-t')}, \\
u(t) &= \theta(t) Q_0 \mathrm{ sin}\left(\omega_0 t\right)e^{-\gamma t} = -\frac{Q_0}{2} D_{R,eq}(t)  .
\end{split}
\end{equation}
where $Q_0$ denotes the  amplitude of the coherent phonon oscillation. We Fourier transform and obtain
\begin{align}
D_{R,1}(t,\omega) &= \frac{ Q_0}{2  \pi}\int D_{R,eq}(\Omega ) D_{R,eq}(\omega - \Omega/2) \nonumber\\
&\times  D_{R,eq}(\omega + \Omega/2)
e^{-i\Omega t} d\Omega,
\end{align}
Finally, we integrate over frequency $\Omega$ we find that
\begin{align}
&D_{R,1}(t,\omega) = Q_0  i e^{-\gamma t} \left[ \frac{4 \omega_0^2 e^{-i\omega_0 t}}{A(\omega,\omega_0)}
- \frac{4 \omega_0^2 e^{i\omega_0 t}}{A(\omega,-\omega_0)}\right]
\nonumber\\
&+
8 Q_0  i e^{- 2 \gamma t} \left[ \frac{4 \omega_0^2 e^{2 i t (\omega + \omega_0 )}}{B(\omega,\omega_0)}
- \frac{4 \omega_0^2 e^{2 i t(\omega-\omega_0 )}}{B(\omega,-\omega_0)}\right]
\end{align}
with
\begin{subequations}
\begin{align}
A(\omega,\omega_0)  &=  (\omega - \frac{3\omega_0}{2}+i\frac{3\gamma}{2})
(\omega + \frac{3\omega_0}{2}+i\frac{\gamma}{2})
\nonumber\\
&\times (\omega + \frac{\omega_0}{2}+i\frac{3\gamma}{2})
(\omega - \frac{\omega_0}{2}+i\frac{\gamma}{2}),\\
B(\omega,\omega_0)  &=   (4 \omega  + 4\omega_0 + 4 i \gamma )( 4\omega + 4 i \gamma )
\nonumber\\
&\times ( 2 \omega + 3 \omega_0 + i\gamma )(2\omega + \omega_0 + i \gamma).
\end{align}
\end{subequations}
From the above expression, we see that the spectral function $A(t, \omega_0)$ evaluated at frequency
$\omega_0$ has both a $4\omega_0$ (third term above) and an $\omega_0$ oscillatory component
(first and second terms above) in time domain.

\section{Spectral function out of equilibrium}
\label{app-c}

In this appendix, we show that the time dependent spectral function can be extracted from the difference
of the Stokes and anti- Stokes Raman intensities. Following the theory of time and frequency resolved
Raman spectroscopy, the scattering intensity $I(t,\omega) \propto D^{>}(t,\omega)$ where $D^{>}(t,\omega)$
is the Wigner transform of the two-time greater function. The relevant references can be found in the main text.
Here, we derive a relation that connects the greater and retarded Green's function out of equilibrium.

The greater component of the Green's function satisfies the relation
\begin{equation}
D^{>}(t,t')= \frac{1}{2}\bigg( D^{K}(t,t')+D^{R}(t,t')-D^{A}(t,t')\bigg).
\end{equation}
We use the linear property of the Wigner transformation and find that
\begin{equation}
D^{>}_{W}(t,\omega)= \frac{1}{2}\bigg( D^{K}_{W}(t,\omega)+D^{R}_{W}(t,\omega)-D^{A}_{W}(t,\omega)\bigg).
\end{equation}
The Keldysh Green's function is symmetric with respect to its two time arguments. Therefore, its
Wigner transform satisfies the relation $ D^{K}_{W}(t,\omega)= D^{K}_{W}(t,-\omega)$. Similarly,
the advanced and retarded Green's function are related by the identity $D^{R}(t,t')=D^{A}(t',t)$,
which implies that $D^{R}_{W}(t,\omega)=D^{A}_{W}(t,-\omega)$.  Thus, we have that
\begin{equation}\label{SQ_fluc_diss_Relation}
\begin{split}
D^{>}_{W}(t,\omega)-D^{>}_{W}(t,-\omega)&= \bigg( D^{R}_{W}(t,\omega)-D^{R}_{W}(t,-\omega)\bigg),\\
&= 2i \, \mathrm{Im}\left[D^{R}_{W}(t,\omega)\right],\\
\end{split}
\end{equation}
where we used $D^{R}_{W}(t,\omega)=D^{R}_{W}(t,-\omega)^{*}$. Finally, we use the fact that the
intensity of the Raman spectra is proportional to the greater Green's function $I(t,\omega) \propto D^{>}(t,\omega)$,
and find that
\begin{equation}
A(t,\omega) \propto I(t,\omega) - I(t,-\omega).
\end{equation}
The above relation shows that  the out of equilibrium spectral function $A(t,\omega)$ can be measured using
time resolved Raman spectroscopy.


\begin{thebibliography}{99}

\bibitem{reviews}
for reviews see, e.g.,
J. Orenstein,
Phys. Today {\bf 65}, 44 (2012);
J. Zhang and R. D. Averitt,
Annu. Rev. Mater. Res. {\bf 44}, 19 (2014);
D. Nicoletti and A. Cavalleri, Adv. Opt. Photon., {\bf 3} 401 (2016);
C. Giannetti, M. Capone, D. Fausti, M. Fabrizio, F. Parmigiani, and D. Mihailovic,
Adv. Phys. {\bf 65}, 58 (2016);
D. N. Basov, R. D. Averitt, D. Hsieh,
Nature Materials, {\bf 16}, 1077 (2017).

\bibitem{okamoto07}
H. Okamoto, H. Matsuzaki, T. Wakabayashi, Y. Takahashi, and T. Hasegawa,
Phys. Rev. Lett. {\bf 98}, 037401 (2007).

\bibitem{fausti11}
D. Fausti, R. I. Tobey, N. Dean, S. Kaiser, A. Dienst, M. C.
Hoffmann, S. Pyon, T. Takayama, H. Takagi, and A. Cavalleri,
Science {\bf 331}, 189 (2011).

\bibitem{liu12}
M. Liu, H. Y. Hwang, H. Tao, A. C. Strikwerda, K. Fan, G. R.
Keiser, A. J. Sternbach, K. G. West, S. Kittiwatanakul, J. Lu,
S. A. Wolf, F. G. Omenetto, X. Zhang, K. A. Nelson, and R. D. Averitt,
Nature {\bf 487}, 345 (2012).

\bibitem{Zong}
A. Zong, P. E. Dolgirev, A. Kogar, E. Erge\c{c}en, M. B. Yilmaz, Ya-Q. Bie, T. Rohwer, I-C. Tung,
J. Straquadine, X. Wang, Y. Yang, X. Shen, R. Li, J. Yang, S. Park, M. C. Hoffmann, B. K. Ofori-Okai,
M. E. Kozina, H. Wen, X. Wang, I. R. Fisher, P. Jarillo-Herrero, and N. Gedik,
Phys. Rev. Lett. {\bf 123}, 097601 (2019).


\bibitem{mitrano16}
M. Mitrano, A. Cantaluppi, D. Nicoletti, S. Kaiser, A. Perucchi, S. Lupi,
P. Di Pietro, D. Pontiroli, M. Ricc\`{o}, S. R. Clark, D. Jaksch, and A. Cavalleri,
Nature {\bf 530}, 461 (2016)

\bibitem{Kogar}
A. Kogar, A. Zong, P. E. Dolgirev, X. Shen, J. Straquadine, Y.-Q. Bie, X. Wang, T. Rohwer,
I.-C. Tung, Y. Yang, R. Li, J. Yang, S. Weathersby, S. Park, M. E. Kozina, E. J. Sie,
H. Wen, P. Jarillo-Herrero, I. R. Fisher, X. Wang, and N. Gedik,
Nat. Phys. {\bf 16}, 159 (2019).

\bibitem{Buzzi}
M. Buzzi, D. Nicoletti, M. Fechner, N. Tancogne-Dejean, M. A. Sentef, A. Georges, M. Dressel,
A. Henderson, T. Siegrist, J. A. Schlueter, K. Miyagawa, K. Kanoda, M.-S. Nam, A. Ardavan,
J. Coulthard, J. Tindall, F. Schlawin, D. Jaksch, and A. Cavalleri,
arXiv:2001.05389.

\bibitem{Li-Eckstein}
J. Li, H. U. R. Strand, P. Werner, and M. Eckstein,
Nat. Commun. {\bf 9}, 4581 (2018).

\bibitem{Matthies-Eckstein}
A. Matthies, J. Li, and M. Eckstein,
Phys. Rev. B {\bf 98}, 180502(R) (2018).

\bibitem{silvestri1985}
S. De Silvestri, J. G. Fujimoto, E. P. Ippen, E. B. Gamble, L. R. Williams, and K. A.
Nelson, Chem. Phys. Lett. {\bf 116}, 146 (1985).

\bibitem{yan1985}
Y.-X. Yan, E. B. Gamble, and K. A. Nelson,
J. Chem. Phys. {\bf 83}, 5391 (1985).

\bibitem{chen1990}
T. K. Cheng, S. D. Brorson, A. S. Kazeroonian, J. S. Moodera, G. Dresselhaus,
M. S. Dresselhaus, and E. P. Ippen,
Appl. Phys. Lett. {\bf 57}, 1004 (1990).

\bibitem{zeiger1992}
H. J. Zeiger, J. Vidal, T. K. Cheng, E.P. Ippen,
G. Dresselhaus, and M. S. Dresselhaus,
Phys. Rev. B {\bf 45}, 768 (1992).

\bibitem{merlin1997}
R. Merlin,
Solid State Commun. {\bf 102}, 207 (1997).

\bibitem{ishioka2009}
for a review see, e.g.,
K. Ishioka, and O. V. Misochko,
Springer Ser. Chem. Phys. {\bf 98}, 23 (2010).

\bibitem{foerst11}
M. F\"{o}rst, C. Manzoni, S. Kaiser, Y. Tomioka, Y. Tokura, R. Merlin and A. Cavalleri,
Nat. Phys. {\bf 7}, 854 (2011).

\bibitem{Subedi}
A. Subedi, A. Cavalleri, A. Georges, Phys. Rev. B {\bf 89}, 220301(R) (2014)

\bibitem{mankowsky14}
R. Mankowsky, A. Subedi, M. F\"{o}rst, S. O. Mariager, M. Chollet, H. T. Lemke,
J. S. Robinson, J. M. Glownia, M. P. Minitti, A. Frano, M. Fechner, N. A. Spaldin,
T. Loew, B. Keimer, A. Georges, and A. Cavalleri,
Nature {\bf 516}, 71 (2014).

\bibitem{lakehal19}
M. Lakehal and I. Paul,
Phys. Rev. B {\bf 99}, 035131 (2019).

\bibitem{hu96}
see, e.g.,
X. Hu and F. Nori,
Phys. Rev. Lett. {\bf 76}, 2294 (1996);
X. Hu and F. Nori, Phys. Rev. B {\bf 53}  2419 (1996)

\bibitem{drummond04}
\emph{Quantum Squeezing}, P.D. Drumond and Z. Ficek (Eds.),
Springer Series on Atomic, Optical, and Plasma Physics (2004).

\bibitem{knap16}
M. Knap, M. Babadi, G. Refael, I. Martin, and E. Demler,
Phys. Rev. B {\bf 94}, 214504 (2016).

\bibitem{kennes17}
D. M. Kennes, E. Y. Wilner, D. R. Reichman, and A. J. Millis,
Nat. Phys. {\bf 13}, 479 (2017).


\bibitem{Wollman}
E. E. Wollman, C. U. Lei, A. J. Weinstein, J. Suh, A. Kronwald, F. Marquardt, A. A. Clerk, K. C. Schwab,
Science {\bf 349} 952 (2015)

\bibitem{Clerk}
A. A. Clerk, K. W. Lehnert, P. Bertet, Y. Nakamura, Nat. Phys. {\bf 16}, 257 (2020)



\bibitem{garrett1997a}
G. A. Garrett, A. G. Rojo, A. K. Sood, J. F. Whitaker, and R. Merlin,
Science {\bf 275}, 1638 (1997).

\bibitem{garrett1997b}
G. A. Garrett, J. F. Whitaker, A. K. Sood, and R. Merlin,
Optics Express {\bf 1}, 387 (1997).

\bibitem{bartels2000}
A. Bartels, T. Dekorsy, and H. Kurz,
Phys. Rev. Lett. {\bf 84}, 2981 (2000).

\bibitem{misochko2000}
O. V. Misochko, K. Sakai, and S. Nakashima,
Phys. Rev. B {\bf 61}, 11225 (2000).

\bibitem{johnson09}
S. L. Johnson, P. Beaud, E. Vorobeva, C. J. Milne, \'{E} .D. Murray, S. Fahy, and G. Ingold,
Phys. Rev. Lett. {\bf 102}, 175503 (2009).

\bibitem{zijlstra10}
E. S. Zijlstra, L. E. D\'{i}az-S\'{a}nchez, and M. E. Garcia,
Phys. Rev. Lett. {\bf 104}, 029601 (2010).

\bibitem{trigo2013}
M. Trigo, M. Fuchs, J. Chen, M. P. Jiang, M. Cammarata, S. Fahy, D. M. Fritz, K. Gaffney,
S. Ghimire, A. Higginbotham, S. L. Johnson, M. E. Kozina, J. Larsson, H. Lemke,
A. M. Lindenberg, G. Ndabashimiye, F. Quirin, K. Sokolowski-Tinten, C. Uher, G.Wang,
J. S.Wark, D. Zhu, and D. A. Reis,
Nat. Phys. {\bf 9}, 790 (2013).

\bibitem{esposito15}
M. Esposito, K. Titimbo, K. Zimmermann, F. Giusti, F. Randi, D. Boschetto, F. Parmigiani, R. Floreanini,
F. Benatti, and D. Fausti,
Nat Comm. {\bf 6}, 10249 (2015).

\bibitem{benatti17}
F. Benatti, M. Esposito, D. Fausti, R. Floreanini, K. Titimbo, and K. Zimmermann,
New J. Phys. {\bf 19}, 023032 (2017).

\bibitem{KamenevBook}
A. Kamenev, \emph{Field Theory of Non-Equilibrium Systems}, Cambridge University Press (2011)

\bibitem{SchiroLeHur}
M. Schir\'o and K. Le Hur, Phys. Rev. B {\bf 89}, 195127 (2014).

\bibitem{supplemental}
see Supplementary Material for technical details.

\bibitem{Wang-Devereaux}
Y. Wang, T. P. Devereaux, and C.-C. Chen,
Phys. Rev. B {\bf 98}, 245106 (2018).

\bibitem{Shvaika}
A. M. Shvaika, O. P. Matveev, T. P. Devereaux, and J. K. Freericks,
Condens. Matter Phys. {\bf 21}, 33707 (2018).

\bibitem{Batignani}
G. Batignani, D. Bossini, N. Di Palo, C. Ferrante, E. Pontecorvo, G. Cerullo, A. Kimel, and T. Scopigno,
Nat. Photonics {\bf 9}, 506 (2015).

\bibitem{Dorfman}
K. E. Dorfman, B. P. Fingerhut, and S. Mukamel,
J. Chem. Phys. {\bf 139}, 124113 (2013).



\end{thebibliography}
\end{document}